\documentclass[iop,apj]{emulateapj}
\usepackage{graphicx}
\usepackage{color}
\usepackage{hyperref}
\usepackage{amsmath}

\def\arcsec{\hbox{$^{\prime\prime}$}}
\def\deg{\hbox{$^\circ$}}
\def\init{\hspace{0.75 mm}}
\def\gala{Was~49a}
\def\galb{Was~49b}
\def\chisq{$\chi^2$}
\def\chisqdof{\chi^2\textrm{/dof}}
\def\nh{N_\textrm{H}}
\def\cmsq{~cm$^{-2}$}
\def\Fxfull{F_{\textrm{0.5-195~keV}}}
\def\Lxfull{L_{\textrm{0.5-195~keV}}}
\def\Fxtwoten{F_{\textrm{2-10~keV}}}
\def\Lxtwoten{L_{\textrm{2-10~keV}}}
\def\FxSwift{F_{\textrm{14-195~keV}}}
\def\LxSwift{L_{\textrm{14-195~keV}}}
\def\ergcms{erg~cm$^{-2}$~s$^{-1}$}
\def\ergs{erg~s$^{-1}$}
\def\nustar{\textit{NuSTAR}}
\def\chandra{\textit{Chandra}}
\def\swift{\textit{Swift}}

\def\xmm{\textit{XMM-Newton}}
\def\oiii{[O\hspace*{1mm}\textsc{iii}]}
\def\nii{[N\hspace{1mm}\textsc{ii}]}
\def\sii{[S\hspace{1mm}\textsc{ii}]}
\def\hei{He\hspace{1mm}\textsc{i}}
\def\caii{Ca\hspace{1mm}\textsc{ii}}
\def\MSol{M$_\sun$}
\def\sersic{S\'{e}rsic}

\definecolor{royalazure}{rgb}{0.0, 0.22, 0.66}
\definecolor{dartmouthgreen}{rgb}{0.05, 0.5, 0.06}

\submitted{}
\begin{document}

\title{Was~49b: An Overmassive AGN in a Merging Dwarf Galaxy?}

\author{Nathan J.\init Secrest\altaffilmark{1,5}, Henrique R.\init Schmitt\altaffilmark{2}, Laura Blecha\altaffilmark{3}, Barry Rothberg\altaffilmark{4,5}, and Jacqueline Fischer\altaffilmark{2}}

\affil{$^1$National Academy of Sciences NRC Research Associate, resident at Naval Research Laboratory; \textcolor{blue}{\href{mailto:nathansecrest@msn.com}{nathansecrest@msn.com}}}

\affil{$^2$Naval Research Laboratory, Remote Sensing Division, 4555 Overlook Ave.\ SW, Washington, DC 20375, USA}

\affil{$^3$University of Maryland Dept.~of Astronomy, 1113 PSC, Bldg.~415, College Park, MD 20742, USA}

\affil{$^4$LBT Observatory, University of Arizona, 933 N.~Cherry Ave., Tuscan, AZ 85721, USA}

\affil{$^5$Department of Physics \& Astronomy, George Mason University, MS 3F3, 4400 University Drive, Fairfax, VA 22030, USA}

\begin{abstract}

We present a combined morphological and X-ray analysis of Was~49, an isolated, dual AGN system notable for the presence of a dominant AGN \galb{} in the disk of the primary galaxy \gala{}, at a projected radial distance of 8~kpc from the nucleus.  Using X-ray data from \chandra{}, \nustar, and \swift, we find that this AGN has a bolometric luminosity of $L_\mathrm{bol}\sim2\times10^{45}$~\ergs{}, with a black hole mass of $M_\mathrm{BH}=1.3^{+2.9}_{-0.9}\times10^8$~\MSol{}.  Despite its large mass, our analysis of optical data from the \textit{Discovery Channel Telescope} shows that the supermassive black hole is hosted by a stellar counterpart with a mass of only $5.6^{+4.9}_{-2.6}\times10^9$~\MSol{}, making the SMBH potentially larger than expected from SMBH-galaxy scaling relations, and the stellar counterpart exhibits a morphology that is consistent with dwarf elliptical galaxies.  Our analysis of the system in the $r$ and $K$ bands indicates that Was~49 is a minor merger, with a mass ratio of \galb{} to \gala{} between $\sim$1:7 and $\sim$1:15.  This is in contrast with findings that the most luminous merger-triggered AGNs are found in major mergers, and that minor mergers predominantly enhance AGN activity in the primary galaxy.  

\end{abstract}

% Online Abstract: We present a combined morphological and X-ray analysis of Was 49, an isolated, dual AGN system notable for the presence of a dominant AGN Was 49b in the disk of the primary galaxy Was 49a, at a projected radial distance of 8 kpc from the nucleus.  Using X-ray data from Chandra, NuSTAR, and Swift, we find that this AGN has a bolometric luminosity of L_bol ~ 2 x 10^45 erg/s, with a black hole mass of M_BH=1.3^{+2.9}_{-0.9} x 10^8 M_Sol.  Despite its large mass, our analysis of optical data from the Discovery Channel Telescope shows that the supermassive black hole is hosted by a stellar counterpart with a mass of only 5.6^{+4.9}_{-2.6} x 10^9 M_Sol, making the SMBH potentially larger than expected from SMBH-galaxy scaling relations, and the stellar counterpart exhibits a morphology that is consistent with dwarf elliptical galaxies.  Our analysis of the system in the r and K bands indicates that Was 49 is a minor merger, with a mass ratio of Was 49a to Was 49b between 1:7 and 1:15.  This is in contrast with findings that the most luminous merger-triggered AGNs are found in major mergers, and that minor mergers predominantly enhance AGN activity in the primary galaxy.  


\section{Introduction}
\label{intro}

Galaxy mergers are generally understood to be major drivers of galaxy evolution, as gravitational torques effectively funnel gas into the central regions of the merging galaxies \citep[e.g.,][]{ToomreToomre72,BarnesHernquist91,DiMatteo+05}.  This inflowing gas enhances star formation \citep[e.g.,][]{Ellison+08,Ellison+10} as well as AGN activity \citep{Koss+10,Ellison+11,Ellison+13,Satyapal+14}.  Along with AGN `feedback' \citep[e.g.,][]{SilkRees98,Springel+05}, these effects have led to a strong correlation between the mass of supermassive black holes (SMBHs) and the mass of their host galaxies' stellar bulge \citep[e.g.,][]{Magorrian+98,Gebhardt+00}, suggesting a coevolution of SMBHs and their host galaxies throughout cosmic history \citep{Richstone+98}.  Empirically, \citet{Ellison+11} find that while major mergers ($M_1$/$M_2>$1/3) enhance AGN activity in both galaxies, minor mergers ($M_1$/$M_2<$1/3) predominantly enhance AGN activity in the larger galaxy during the merger. Similarly, \citet{Koss+12} find that the most powerful AGNs in minor mergers are triggered in the more massive galaxy.

In this context, the dual AGN system Was~49 \citep{Bothun+89} is quite peculiar.  The system is composed of a disk galaxy, \gala{}, hosting a low luminosity Seyfert 2 nucleus, and a powerful Type 2 AGN, \galb{}, co-rotating within the disk \citep{Moran+92} at a projected distance of $\sim8$~kpc from the center of \gala{}.  \citet{NishiuraTaniguchi98}, using the broad ($\sim6000$~km~s$^{-1}$) polarized H$\beta$ emission seen in \galb{}, determine a black hole mass of $2.9\times10^8$~\MSol.  The optical continuum is almost featureless, and previous estimates led to an upper limit on any stellar component in the optical continuum of $\lesssim{15\%}$ \citep{Tran95a}.  Since these studies were published, the discovery of SMBH/galaxy scaling relations has provided a new context for the unusual nature of \galb{}, with its large SMBH and apparent lack of a stellar counterpart.

In this paper, we quantify the peculiarities of \galb{}.  We determine the intrinsic luminosity and accretion rate of the AGN through a detailed X-ray spectral analysis, and we combine the results of our X-ray analysis with an analysis of its optical spectrum to derive an independent measure of the black hole mass.  We perform a morphological analysis of the Was~49 system to estimate the mass of the stellar counterpart to \galb{}, and we estimate the stellar mass ratio of the primary galaxy \gala{} to \galb{}.  We adopt $H_0=70$~km~s$^{-1}$~Mpc$^{-1}$, $\Omega_{\textrm{M}}=0.3$, and $\Omega_\Lambda=0.7$.  We use the redshift to the nucleus of the primary galaxy Was~49a ($z=0.06328$) for distances.

% NED: 	luminosity distance = 288~Mpc
%		angular size distance = 255 Mpc
%		comoving radial/tang distance = 271 Mpc
%		scale = 1.235 kpc/arcsec

\section{Methodology}

\subsection{X-ray Analysis}
\label{xraydata}
In order to constrain the X-ray spectral parameters of \galb{} as tightly as possible, we obtained all archival X-ray data suitable for spectral fitting.  These datasets come from the \chandra{}~X-ray~Observatory, the \textit{Nuclear~Spectroscopic~Telescope~Array} \citep[\nustar{};][]{Harrison+13}, and the \swift{} Burst~Alert~Telescope \citep[BAT;][]{Krimm+13}.  There is also data from the \textit{Advanced Satellite for Cosmology and Astrophysics} \citep[ASCA;][]{TanakaInoue94}, however an estimated 10\% of this low spatial resolution data is contaminated by RX~J1214.4+2936, located 4\arcmin\hspace{-0.9mm}.6 away \citep[see][]{Awaki+00}, so we elect not to use it. 

% SWIFT DATA
\vspace{0.1cm} \noindent \underline{\textbf{\swift{}}:} We downloaded the BAT spectrum and response file directly from the BAT 70-month Hard X-ray Survey \citep{Baumgartner+13} webpage.\footnote{\label{70mon}\url{http://swift.gsfc.nasa.gov/results/bs70mon/}}  The BAT spectrum covers the period from 16~December~2004 to 30~September~2010, and has an effective exposure time of 13.1~Ms, yielding a total signal-to-noise (S/N) of 8.64.

% NUSTAR DATA
\vspace{0.1cm} \noindent \underline{\textbf{\nustar{}}:} An archival \textit{NuSTAR} observation (ObsID=60061335002) of \galb{} was taken 15~July~2014 as part of the extragalactic survey \citep{Harrison+15}.  We created Level 2 event files by running \texttt{nupipeline}, version 0.4.4, with the latest \textit{NuSTAR} CALDB files.  We created source and background region files in ds9 by using two 30$\arcsec$ radius circular apertures, one at the position of \galb{}, and the other on the same detector between the source and the edge of the image. We created Level 3 data products for the FPMA and FPMB data, including source and background spectra, using \texttt{nuproducts}, and setting the \texttt{rungrppha} flag to produce grouped output spectra.  We required a minimum grouping of 30 counts for the \chisq{} statistic, and we set the high and low bad data energy channel thresholds at 1909 and 35, corresponding to 3-78~keV.  Because \nustar{} has two detectors, inter-instrumental sensitivity differences may affect our analysis.  To address this, we fit the FPMA and FPMB data with a single absorbed power-law model between 3-78~keV, appending a constant to each detector group, and holding the FPMA constant fixed at unity.  We find the FPMB constant equal to $1.07^{+0.10}_{-0.90}$, consistent with a negligible instrumental sensitivity difference.

\begin{figure}
\noindent{\includegraphics[width=\columnwidth]{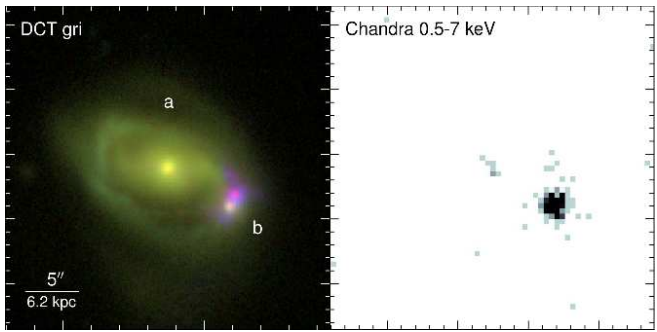}}
\caption{Optical and X-ray images of the Was~49 system, with optical color representation following \citet{Lupton+04}.  At the redshift of Was~49 ($z=0.06328$), \oiii{} and H$\beta$ fall within the $g$ (blue) filter, the $r$ (green) filter is predominantly continuum, and the $i$ (red) filter contains H$\alpha+$\nii{}.  Note the extensive ionization region around \galb{}, indicating that the optical light is dominated by line emission. Images oriented north up, east left.}
\label{fig:griX}
\end{figure}

% CHANDRA DATA
\vspace{0.1cm} \noindent  \underline{\textbf{\chandra{}:}} A 5~ks ACIS-I dataset (ObsID=14042) exists for \galb{}, taken on 2012~March~25.  We reprocessed the event and calibration files for this dataset using \textsc{ciao}, version 4.7, and CALDB~4.6.9.  We limited our reprocessed event files to events between 0.5-7~keV, and we used \texttt{dmextract} with a 3$\arcsec$ radius aperture and a nearby background region to extract counts.  We give a summary of the \chandra{} data for the Was~49 system in Table~\ref{tab:ChandraSources}, but we note that we find a 7 count (2.6$\sigma$) detection at the nucleus of the larger galaxy Was~49a, which may be in line with earlier findings that this is a dual AGN system.  We show the \chandra{} image of the Was~49 system in Figure~\ref{fig:griX}, right.

\begin{deluxetable}{lccrr}
\tablecaption{\chandra{} Information \label{tab:ChandraSources}}
\tablehead{
\colhead{}			& \colhead{}	& \colhead{}		& \colhead{counts}		& \colhead{counts}  	\\ [-0.1cm]
\colhead{Was~49}	& \colhead{R.A.}	& \colhead{Decl.}	& \colhead{}			& \colhead{}		\\ [-0.1cm]
\colhead{}			& \colhead{}	& \colhead{}		& \colhead{0.5-7~keV}	& \colhead{2-7~keV}	}
\startdata
a	& 12\textsuperscript{h}14\textsuperscript{m}18\textsuperscript{s}\hspace{-0.5mm}.262	& +29$\deg$31$\arcmin$46$\arcsec$\hspace{-0.9mm}.75		& 7$^*$		& 0		\\ 
b	& 12\textsuperscript{h}14\textsuperscript{m}17\textsuperscript{s}\hspace{-0.5mm}.816	& +29$\deg$31$\arcmin$43$\arcsec$\hspace{-0.9mm}.17		& 215	& 159
\enddata
\tablenotetext{$*$}{The 1$\sigma$ statistical uncertainty on the source position for a 7-count source, estimated using \texttt{wavdetect}, is $\sim0\arcsec\hspace{-0.9mm}.7$.}
\end{deluxetable}

We masked all the sources in the event data using regions obtained with \texttt{wavdetect}, and we extracted the background light curve using \texttt{dmextract}.  We found no significant ($>3\sigma$) flaring intervals during this 5~ks observation.  Within the same 30$\arcsec$ aperture that we used for the \nustar{} data is a third X-ray source, in addition to the nuclear source in Was~49a, without any obvious association with the Was~49 system, with 9 counts between 0.5-7~keV.  Adding this to the Was~49a nuclear source, we find a contamination level within the \nustar{} aperture of $(7^{+3}_{-2})\%$ between 0.5-7~keV, where the uncertainty represents the 90\% confidence interval calculated using Poisson statistics.  However, by extracting counts only in the harder 2-7~keV band, the contamination level drops to $(3\pm2)\%$.  We therefore do not attempt to make any sort of contamination correction to our X-ray spectral analysis, which is dominated by the hard X-ray data from \textit{NuSTAR} and \textit{Swift}.  We extracted X-ray spectra using \texttt{specextract}, and we used the same 3$\arcsec$ aperture we used to isolate emission from \galb{}, as we seek to minimize contamination at soft energies by other sources.  Using \textsc{grppha}, we grouped the spectra by a minimum of 20 counts for the \chisq{} statistic.

We performed spectral analyses on the grouped spectra\footnote{with the exception of the BAT spectrum, as it does not require grouping; see \url{http://swift.gsfc.nasa.gov/analysis/threads/batspectrumthread.html}} using \textsc{xspec}, version 12.9.0~\citep{arnaud1996}. We confine our X-ray spectral analysis to 0.5-7~keV, 3-78~keV, and 14-195~keV for the \chandra{}, \nustar{}, and \swift{} spectra, respectively. To derive fluxes and uncertainties for specific model components, we appended the \texttt{cflux} convolution model, holding the additive model component of interest normalization fixed.  We model the Galactic hydrogen column density as a fixed photoelectric absorption multiplicative component (\texttt{phabs}) with a value of $\nh{}=1.88\times10^{20}$\cmsq{}, calculated using the \swift{} \texttt{nhtot} tool\footnote{\url{http://www.swift.ac.uk/analysis/nhtot/}}, which uses the prescription of \citet{Willingale+13}.  All X-ray spectral parameters, fluxes, and uncertainties given in this work are the values that minimize the \chisq{} statistic, along with their 90\% confidence intervals.

\subsubsection{Inter-Epoch Variability}
\label{OffsetsAndVariability}

Owing to the difference in epochs between the X-ray data, variability may hamper joint spectral fitting.  This is especially a concern for the higher-energy \nustar{} and \swift{} datasets, since the highest-energy photons are produced in the innermost accretion regions of the AGN.  To address this issue, we looked for variability in the \chandra{}, and \swift{} data relative to the \nustar{} data, because \nustar{} shares its energy range with the other two.  We first fit the \chandra{} and \nustar{} data with an absorbed power-law model, allowing the model normalizations to vary.  We find the \chandra{} and \nustar{} normalizations (keV$^{-1}$~cm$^{-2}$~s$^{-1}$) to be $6.2^{+3.4}_{-2.1}\times10^{-4}$ and $7.6^{+3.5}_{-2.3}\times10^{-4}$, respectively, indicating no discernible variability between the two epochs.

We separately find normalizations of $8.4^{+7.2}_{-3.8}\times10^{-4}$ and $8.5^{+4.5}_{-2.8}\times10^{-4}$ for \swift{} and \nustar{}, respectively, making any flux variability between the datasets below the threshold of detectability.  We note that in the above analysis, we have implicitly assumed that the power-law spectral index $\Gamma$ has remained unchanged from epoch to epoch.  \citet{Soldi+14} find, using data from the BAT 58-month survey, that the majority of AGNs in the survey do not exhibit significant spectral variability. \citet{Hernandez-Garcia+15} similarly find, using a sample of 25 Seyfert~2 galaxies with \chandra{}/\xmm{} observations, that X-ray variability is primarily in flux (e.g., the normalization of their power-law spectra) and not spectral shape (e.g., $\Gamma$).  Moreover, setting the normalizations of the power-law fit to the \swift{} and \nustar{} data equal and allowing the spectral indices $\Gamma$ to vary, we find $\Gamma=1.6\pm0.1$ and $\Gamma=1.6^{+0.2}_{-0.1}$ for \swift{} and \nustar{}, respectively, consistent with insignificant spectral variability.

\subsection{Estimating the Black Hole Mass}
\label{subsec:methodBHmass}

To obtain an independent measure of the mass of the black hole in \galb{}, we used the optical spectrum of \galb{} from the Baryon Oscillation Spectroscopic Survey \citep[BOSS;][]{Dawson+13} to determine the FWHM of the broad H$\alpha$ line, and we used the intrinsic X-ray luminosity (\S\ref{X-rayResults}) to estimate the size of the broad line region (BLR) using the $\Lxtwoten$-$R_\mathrm{BLR}$ relation from \citet{Kaspi+05}.  We use the intrinsic X-ray luminosity, instead of the $\lambda L_\lambda$(5100~\AA) luminosity commonly used to estimate the size of the BLR \citep[e.g.,][]{Kaspi+00}, because \galb{} is a known polarized BLR AGN \citep[e.g.,][]{Tran+92,Tran95b}, and so most if not all of the optical continuum is scattered to some degree, making it difficult to determine the intrinsic $5100$~\AA{} luminosity.  Moreover, the relation we employ here has been shown to be in agreement with maser-determined black hole masses for a sample of four polarized BLR AGNs, which includes NGC~1068 and Circinus \citep{Kuo+11}.

The optical continuum of \galb{} is almost entirely non-stellar, so we effectively subtracted the continuum by fitting a simple power-law model, after masking the emission lines.  We treated the residuals after continuum subtraction as an additional source of uncertainty by adding the standard deviation of the residuals near lines of interest to the square root of the spectral variance, in quadrature.

The emission lines of \galb{} exhibit a complex morphology, showing multiple broadened components in the forbidden lines.  To prevent this complex morphology from affecting our measure of the FWHM of H$\alpha$, we first modeled the \sii{} $\lambda\lambda$6713, 6731 doublet, which provides an accurate parameterization of the non-BLR line emission \citep[e.g.,][]{FilippenkoSargent88,Ho+97}, as a combination of several Gaussian components, with the wavelength of each member of the doublet separated by the laboratory difference, and allowing each Gaussian component to vary in redshift as well as FWHM and flux.  Once a suitable model of the \sii{} doublet was found, we applied this model to the H$\alpha$+\nii{} complex, holding the \nii{} $\lambda6583/\lambda6548$ flux ratio equal to the theoretical value of 2.96, and setting their wavelengths to the laboratory difference.  We then added an additional broad line component to H$\alpha$, and we also include the \hei{} $\lambda6678$ line seen in the spectrum, as it overlaps with the extended broad H$\alpha$ emission.  The free parameters we fit using this model are the H$\alpha$, \nii{} $\lambda6583$, and \hei{} fluxes, and the FWHM and redshift of the additional broad H$\alpha$ line.  In both the fit to the \sii{} doublet and the fit to the H$\alpha$+\nii{} complex, we used the SciPy non-linear least squares fitting routine \texttt{curve\_fit}, weighting by the uncertainty of the spectrum, and correcting for the wavelength-dependent instrumental spectral resolution.

\subsection{Morphological Analysis}

We observed Was~49 on 3~April~2016 with the \textit{Discovery Channel Telescope} (DCT) Large~Monolithic~Imager, in the Sloan $u^\prime g^\prime r^\prime i^\prime z^\prime $ bands.  Owing to exceptional seeing ($\sim0\arcsec$\hspace{-0.9mm}.5) during the observing run, we were able to achieve a much sharper view of Was~49 than is available from the SDSS archival images.  We obtained $5\times100$~second images in each filter, dithering by 30$\arcsec$ between images.  For the $u^\prime$ band images, we took sky flats, and we took dome flats for the $g^\prime$, $r^\prime$, and $i^\prime$ band.  The fringing typically seen in the $z^\prime$ band has a strong dependence on illumination, so we used the $z^\prime$ band images themselves to make a flat field by taking their median.  The airmass of the observations was between 1.01-1.06.

After reducing our data using the flat and bias frames, we removed cosmic rays using the Python implementation of L.A.Cosmic \citep{vanDokkum01}\footnote{\url{http://obswww.unige.ch/~tewes/cosmics_dot_py/}}.  We found that the sky background in our data is very uniform, so we chose a number of source-free regions neighboring Was~49 to calculate the mean sky background, which we subtracted from our images.  We aligned our images using the \textsc{iraf} task \texttt{imalign}.  We co-added the aligned images within each band to produce mean and variance images.  Our final, co-added images have an angular resolution of $\sim0\arcsec\hspace{-0.9mm}.5-0\arcsec\hspace{-0.9mm}.6$ (FWHM).

We initially corrected the astrometric solution of our data with \textit{Astrometry.net}, version 0.67 \citep{Lang+10}, using SExtractor, version 2.19.5 \citep{BertinArnouts96} and a custom index file we built directly from the SDSS~DR12 \texttt{PhotoObj} table.  We then refined our astrometric solution in the following manner: first, we cross-matched sources in the \texttt{PhotoObj} table with the \texttt{SpecObj} table to produce a list of extragalactic sources in our field.  We then randomly sampled between 20 and 80 sources in the \texttt{PhotoObj} table in an iterative manner, at each iteration performing a least squares fit to a linear six parameter plate model and using the resultant WCS solution to cross-match the position of the extragalactic sources in our field to their SDSS positions.  If the mean value of the astrometric residuals is improved at a given iteration, we stored the new WCS solution.  We were able to achieve a mean astrometric accuracy \textit{relative} to SDSS of 24~milliarcseconds with 68 reference sources.

We calculated photometric zeropoints by using SExtractor to obtain instrumental magnitudes and fluxes, which we fit to the SDSS \texttt{PhotoObj} catalog fluxes by using unresolved sources in our data and comparing their instrumental magnitudes to their point spread function (PSF) magnitudes.  We fit the instrumental magnitudes to their PSF magnitudes by selecting all sources with \texttt{psfMag}~$<21$ to minimize uncertainty due to low S/N.  As expected, the instrumental and PSF magnitudes of these sources are tightly correlated, with a residual scatter of 0.01-0.02 mag.  We hereafter refer to the DCT Sloan magnitudes as simply $ugriz$.  We show the flux-calibrated DCT $gri$ image of Was~49 in Figure~\ref{fig:griX}, left.

% u-band image shift errors: 0.030-0.036 pix
% g-band image shift errors: 0.014-0.016 pix
% r-band image shift errors: 0.010-0.012 pix
% i-band image shift errors: 0.009-0.014 pix
% z-band image shift errors: 0.008-0.011 pix

We fit the Was~49 system with \textsc{galfit}, version 3.0.5 \citep{Peng+02,Peng+10}, using the $r$-band image.  This band was selected because it does not overlap with major emission lines (H$\beta$, \oiii{}, H$\alpha$+\nii{}) at the redshift of the system, and a comparison with the BOSS optical spectrum of \galb{} indicates that the $r$-band has a minimum of emission line contamination ($\sim6\%$).  The $r$ variance image as described above was used as weighting and to calculate the reduced chi-squared of the fit, and our model of the system was convolved with an empirical PSF template constructed from stars within a few arcminutes of Was~49.

% ===================================================================================================
% RESULTS
% ===================================================================================================

\section{Results}
\label{results}

\subsection{X-rays}
\label{X-rayResults}
\begin{figure}
\noindent{\includegraphics[width=\columnwidth]{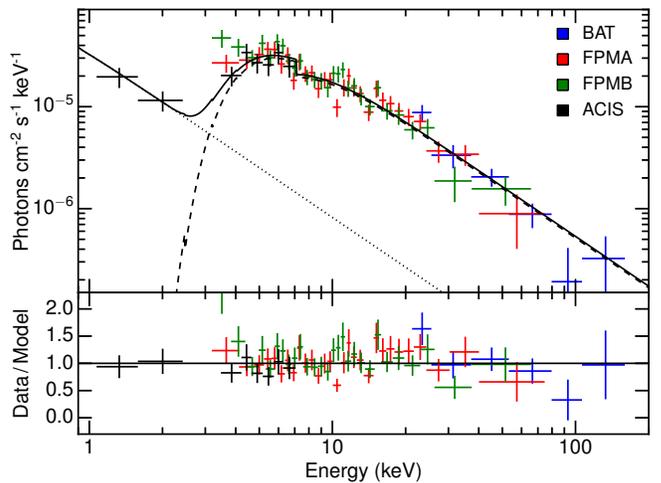}}
\caption{Unfolded rest frame X-ray spectrum of \galb{} with the best fit model.  The absorbed power-law component is shown as the dotted line, and the scattered component is the dashed line.}
\label{fig:spectrum}
\end{figure}

We fit the X-ray spectrum of \galb{} with an absorbed power-law model.  To account for the fraction of the intrinsic continuum that is scattered at large radii \citep[e.g.,][]{Awaki+00}, we append an additional power-law component multiplied by a constant that is free to vary, representing the scattered fraction. Explicitly, our model is: \texttt{phabs*(zphabs*zpow+const*zpow)}, with the second power-law component's parameters tied to the first.  We find a good fit, with $\chisqdof{}=81.76/67$, $\Gamma=1.6\pm0.1$, $\nh{}=2.3_{-0.4}^{+0.5}\times10^{23}$\cmsq{}, and a scattering fraction of $3.8^{+1.6}_{-1.2}\%$ (see Figure~\ref{fig:spectrum}).  Our model yields unabsorbed fluxes (\ergcms{}) of $\Fxfull=(2.4\pm0.2)\times10^{-11}$, $\Fxtwoten=4.0_{-0.6}^{+0.7}\times10^{-12}$, and $\FxSwift=1.7_{-0.2}^{+0.3}\times10^{-11}$.  The corresponding intrinsic X-ray luminosities (\ergs{}) are $\Lxfull=(2.4\pm0.2)\times10^{44}$, $\Lxtwoten=4.0_{-0.6}^{+0.7}\times10^{43}$, and $\LxSwift=1.7_{-0.2}^{+0.3}\times10^{44}$.  We note an apparent absence of the Fe~K$\alpha$ 6.4~keV in the spectrum of \galb{}.  We calculated an upper limit on the strength of this line by freezing the model parameters and adding a Gaussian model component at 6.4~keV with a width of $\sigma=0.1$~keV.  We calculate a 90\% upper limit on the equivalent width (EW) of this line of $\sim0.08$~keV.  To test for the effect of grouping on our data, we repeated this procedure for an ungrouped version of our data, using Cash statistics \citep{Cash79}, but we found nearly identical results, with a 90\% upper limit on the EW of $\sim0.07$~eV.  This upper limit on the Fe~K$\alpha$ EW, however, is not inconsistent with expectations for hydrogen column densities of $\sim10^{23}$\cmsq{} \citep[e.g.,][]{MurphyYaqoob09, BrightmanNandra11}.  To estimate the bolometric luminosity and accretion rate of \galb{}, we used the relation between $L_\mathrm{bol}$ and $\LxSwift{}$ from \citet{Winter+12}, giving $L_\mathrm{bol}\sim1.3\times10^{45}$~\ergs{}. Given $L_\textrm{bol} = \eta\dot{M}c^2$ and assuming a typical accretion efficiency $\eta=0.1$, $\dot{M}\sim0.2$~\MSol~yr$^{-1}$.
 
For \gala{}, the non-detection of X-ray counts above $>2$~keV (Table~\ref{tab:ChandraSources}) implies that we should not assume a typical AGN power law X-ray spectrum.  Indeed, using the Bayesian~Estimation~of~Hardness~Ratios code \citep{Park+06}, we find that the 90\% statistical upper limit on the hardness ratio, defined as (H-S)/(H+S), is -0.9, requiring a power law index greater than $\gtrsim4$, suggesting significant contamination by soft X-ray photons of a non-AGN origin.  A detailed discussion of the nature of the X-ray source in \gala{} is beyond the scope of this work, and we defer it to when deeper X-ray data become available.

\subsection{Black Hole Mass}
\label{subsec:BHmass}
\begin{figure}
\noindent{\includegraphics[width=\columnwidth]{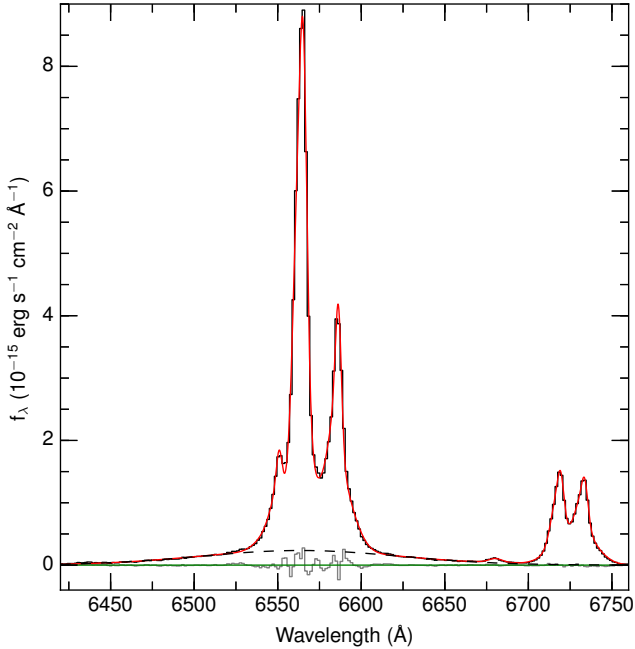}}
\caption{BOSS spectrum of \galb{}.  The model spectrum is in red, the residuals are in grey, and the zero flux line is in green.  We plot the additional broad H$\alpha$ component as the dashed black line.}
\label{fig:Halpha}
\end{figure}

Using a model of three Gaussian components derived from the \sii{} doublet, we achieved a good fit to the H$\alpha$+\nii{} complex after adding the additional broad line component (Figure~\ref{fig:Halpha}).\footnote{Because of the weakness of the \hei{} line, one of the three Gaussian components had a flux consistent with zero.  We removed this component to better constrain our model.}  The FWHM of this broad line component is $6440\pm60$~km~s$^{-1}$, where the uncertainty is 1$\sigma$ and derived from the fit covariance matrix.   As a check on this uncertainty, we created a `null' spectrum without any additional broad H$\alpha$ component, and using the best-fit parameters for the rest of the line components of H$\alpha$, \nii{}, and \hei{}.  We then made $10^4$ permutations of this spectrum, adding to each a broad line with a FWHM randomly chosen between 4000 and 7000 km~s$^{-1}$, and with a flux randomly chosen between 1/3 and 3 times the flux of the line measured in the original spectrum.  We then added Gaussian noise to the spectrum with sigma taken from the spectral uncertainty described in \S\ref{subsec:methodBHmass}.   After fitting each spectrum with our model, we find that the scatter between the input FWHM and the fit FWHM has a standard deviation of $\sim100$~km~s$^{-1}$, which we take as a more realistic estimate of the uncertainty.  We repeated this procedure on the SDSS spectrum of \gala{}, but we did not find evidence for an additional broad line component. 

Additionally, we explored any possible effect of the use of Gaussian components to parameterize the \sii{} doublet by repeating our line fitting procedure using Lorentzian components.  We tested a model with and without a broad line component, and compared their \chisq{} values.  The addition of a Gaussian broad line to the Lorentzian model reduces \chisq{} by a factor of $\sim7$, with FWHM~=~6540~km~s$^{-1}$, consistent with the Gaussian-based \sii{} model within the uncertainties.  However, the model using Gaussian components yields a much better overall fit to the data, with a \chisq{} that is 42\% that of the Lorentzian model, and a KS test shows that the residuals of the Gaussian model are consistent with the spectral uncertainties ($p=0.18$), while the residuals of the Lorentzian model are not ($p=8.2\times10^{-5}$).  We therefore use the FWHM of the broad line from the Gaussian-based model, not only because of the better fit but also because it yields a more conservative estimate of the FWHM of the broad line.

We note that the BOSS spectrum aperture overlaps somewhat with a knot of H$\alpha$-heavy emission seen in Figure~\ref{fig:rgiig}, top and bottom left panels.  While the method of modeling the non-BLR emission we employ is designed to avoid uncertainties with respect to the source of broad emission, it is nonetheless helpful to consider how significantly a knot of emission outside the AGN in \galb{} might be affecting our measure of the FWHM of the BLR.  The broadest component of our \sii{} model is indeed consistent with highly broadened emission in the environment around the AGN, with a FWHM of $\sim960$~km~s$^{-1}$, so a highly broadened non-BLR emission component is already factored into our fit of the H$\alpha+$\nii{} complex.  However, we tested for the possibility that the \nii{} and non-BLR H$\alpha$ line profiles might deviate significantly from \sii{} by freeing their widths and velocities and refitting the spectrum.  Freeing these parameters improved the fit, as expected, but the effect on the FWHM of the BLR component was not significant, increasing it to 6790~km~s$^{-1}$, with about a factor of 2 increase in the uncertainty.  To remain conservative, we retain our estimate of 6440~km~s$^{-1}$ for the FWHM of the BLR, and we adopt a factor of 2 increase of the uncertainty, equal to the uncertainty of 200~km~s$^{-1}$ assumed by \citet{Kuo+11}.

We calculate the black hole mass as $M_\mathrm{BH} = fR_\mathrm{BLR}\sigma_\mathrm{H\alpha}^2/\mathrm{G}$, where we adopt a virial coefficient of $\log{(f)}=0.72$ \citep{Woo+10}, which has an intrinsic scatter of 0.44~dex, and we calculate $R_\mathrm{BLR}\sim18$~light-days with an uncertainty of 52\% \citep{Kaspi+05}.  Propagating these uncertainties and the uncertainty of the FWHM of the broad H$\alpha$, $\log{(M_\mathrm{BH}/\textrm{\MSol{})}}=8.1\pm0.5$, where the uncertainty is 1$\sigma$, consistent with the mass estimate from \citet{NishiuraTaniguchi98}. 

Finally, we note that while the $\Lxtwoten$-$R_\mathrm{BLR}$ relation from \citet{Kaspi+05} that we use has been shown to predict maser masses for several highly obscured AGNs like \galb{} to a high degree of accuracy \citep{Kuo+11}, there are some possible caveats.  The objects studied in \citet{Kuo+11} had typical $\Lxtwoten$ luminosities of about $10^{42}$~\ergs{}, over an order of magnitude less luminous than \galb{}.  Of the objects studied in \citet{Kaspi+05} with $\Lxtwoten$ luminosities similar to \galb{}, while most also have BLR sizes of around $\approx20$ light-days, there are some outliers.  The most notable is IC~4392A, with a BLR size of $\sim2$~light-days, despite having $\Lxtwoten=4.5\times10^{43}$~\ergs{}.  If \galb{} has a BLR of similar size, then its SMBH has a mass of $\log{(M_\mathrm{BH}/\textrm{\MSol{})}}\sim7.2$, and it is radiating very close to its Eddington limit.  This is inconsistent with its X-ray spectral properties, as we discuss in \S\ref{sec:Discussion}, but moreover the BLR size uncertainty for IC~4392A is very large, and as such it is not included in other works on the radius-luminosity relationship \citep[e.g.,][]{Bentz+09}.  Conversely, other objects with similar $\Lxtwoten$ luminosities to \galb{} that have much smaller BLR size uncertainties have BLR sizes that are quite large, up to $\sim100$~light-days.  This being the case, the SMBH in \galb{} has a mass of $\log{(M_\mathrm{BH}/\textrm{\MSol{})}}\sim8.9$.  Neither of these extremes is inconsistent with the 0.5~dex uncertainty that we have derived for our SMBH mass, and the SMBH may in fact be somewhat larger given the uncertainty associated with IC~4392A.  With these considerations, we use the value of $\log{(M_\mathrm{BH}/\textrm{\MSol{})}}=8.1\pm0.5$ originally calculated, noting the caveats detailed above.

\subsection{Morphology}
\label{subsec:Morphology}

\begin{figure}
\noindent{\includegraphics[width=\columnwidth]{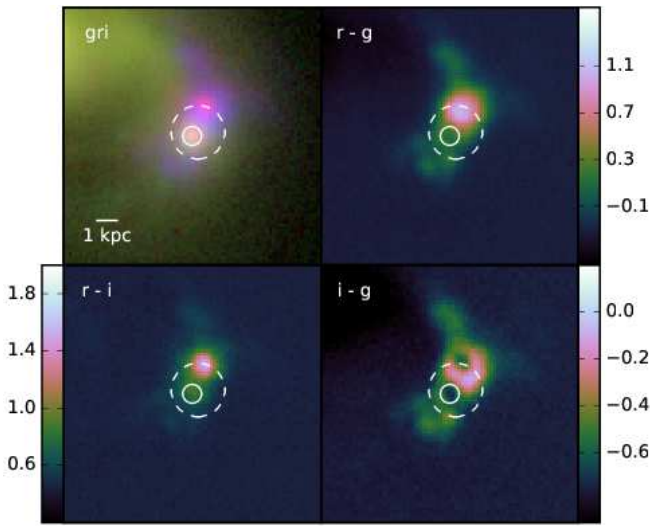}}
\caption{\textit{Top left:} Cutout of Figure~\ref{fig:griX} with the position of the 2$\arcsec$ BOSS spectrum aperture overlaid as the dashed white circle, and the position of the \chandra{} X-ray source overlaid as the solid white circle.  In the case of the latter, the radius denotes the $\sim90\%$ confidence level on the source position. \textit{Top right:} As the $r$ band is mostly uncontaminated by strong line emission, while the $g$ band is strongly contaminated by \oiii{} and H$\beta$ emission (56\% of the light integrated over the filter pass), the $r-g$ traces the observed strength of \oiii{}+H$\beta$.  \textit{Bottom left:} $r-i$ traces the observed strength of H$\alpha$+\nii{}, given the strong contamination of H$\alpha$ and \nii{} in the $i$ band.  \textit{Bottom right:} The $i-g$ image shows the observed strength of \oiii{} relative to H$\alpha$+\nii{}, suggesting that the hard AGN radiation field of \galb{} extends out to several  kpc from the SMBH.  Interestingly there is a minimum of $i-g$ at the position of the AGN itself (as indicated by the location of the \chandra{} source), indicating that the region may be dominated by a stellar population and have a smaller contribution from line emission.}
\label{fig:rgiig}
\end{figure}

The best fit morphological model consists of seven model components, presented in Table~\ref{tab:galfit}. \gala{} was fit with four components, and \galb{} was fit with three.  Both \gala{} and \galb{} have an unresolved nuclear source, although the source in \gala{} is considerably fainter.  The unresolved nuclear source in \galb{} is coincident with the \chandra{} X-ray source and is buried within a region of extensive ionization (Figure~\ref{fig:rgiig}, top left panel).\footnote{By registering background sources in the \chandra{} data with their optical counterparts in the DCT images, we estimate that the 90\% confidence radius of the X-ray source position is $\sim0\arcsec\hspace{-0.9mm}.4$.}  The surrounding ionized region extends $\sim1\arcsec\hspace{-0.9mm}.5$ (1.9~kpc) to the north west and the south east of the stellar concentration, and appears to be stratified into knots and filamentary structures of \oiii{} and H$\alpha$+\nii{} emission (Figure~\ref{fig:rgiig}, remaining panels).

\begin{deluxetable*}{lrrrrrrrrr}
\centering
\tablecaption{\textsc{galfit} Morphological Parameters \label{tab:galfit}}
\tablehead{
\colhead{}			&                        &         		     &\colhead{$M_r^*$}	& \colhead{$\log{L_r}^\dagger$}	& \colhead{R$_\textrm{eff}$}	& \colhead{R$_\textrm{eff}$}		&  \colhead{}				& \colhead{}	& \colhead{PA}		\\ [-0.1cm]
\colhead{Was~49}	& \colhead{RA}  & \colhead{Decl.} & \colhead{}				& 					&\colhead{}				& \colhead{}			& \colhead{\sersic{} $n$}		& \colhead{b/a}	&\colhead{}		\\ [-0.1cm]
\colhead{}			&                       &        		    & \colhead{mag}			& \colhead{[L$_\sun$]}	& \colhead{$\arcsec$} 		& \colhead{kpc}			& \colhead{}				& \colhead{}	& \colhead{deg}	}
\startdata
a: nucleus & 12\textsuperscript{h}14\textsuperscript{m}18\textsuperscript{s}\hspace{-0.5mm}.256 & +29$\deg$31$\arcmin$46$\arcsec$\hspace{-0.9mm}.66 & -16.7 & 8.54 & ... & ... & ... & ... & ... \\ 
a: bulge & 12\textsuperscript{h}14\textsuperscript{m}18\textsuperscript{s}\hspace{-0.5mm}.256 & +29$\deg$31$\arcmin$46$\arcsec$\hspace{-0.9mm}.66 & -20.0 & 9.86 & 0.96 & 1.19 & 1.43 & 0.80 & 82.9 \\ 
a: disk & 12\textsuperscript{h}14\textsuperscript{m}18\textsuperscript{s}\hspace{-0.5mm}.281 & +29$\deg$31$\arcmin$47$\arcsec$\hspace{-0.9mm}.12 & -20.8 & 10.2 & 5.26 & 6.49 & 0.42 & 0.41 & 77.2 \\ 
a: tidal & 12\textsuperscript{h}14\textsuperscript{m}18\textsuperscript{s}\hspace{-0.5mm}.303 & +29$\deg$31$\arcmin$44$\arcsec$\hspace{-0.9mm}.86 & -21.0 & 10.3 & 8.02 & 9.90 & 0.36 & 0.73 & 27.1 \\ 
b: nucleus & 12\textsuperscript{h}14\textsuperscript{m}17\textsuperscript{s}\hspace{-0.5mm}.819 & +29$\deg$31$\arcmin$43$\arcsec$\hspace{-0.9mm}.11 & -17.8 & 8.99 & ... & ... & ... & ... & ... \\ 
b: bulge & 12\textsuperscript{h}14\textsuperscript{m}17\textsuperscript{s}\hspace{-0.5mm}.816 & +29$\deg$31$\arcmin$43$\arcsec$\hspace{-0.9mm}.17 & -19.1 & 9.50 & 1.31 & 1.62 & 1.07 & 0.68 & 344.0 \\ 
b: ionization & 12\textsuperscript{h}14\textsuperscript{m}17\textsuperscript{s}\hspace{-0.5mm}.789 & +29$\deg$31$\arcmin$43$\arcsec$\hspace{-0.9mm}.79 & -18.2 & 9.14 & 0.52 & 0.64 & 0.41 & 0.63 & 340.1
\enddata
\tablenotetext{}{Note: the labeling of components by `a' or `b' is based on their spatial association.\\
$^*$ Calculated using a distance modulus = 37.3 and without a K-correction, which is 0.04 for a color of $g-r=0.5$ at the redshift of the Was~49 system, using the K-corrections calculator at \url{http://kcor.sai.msu.ru/}.\\
$^{\dagger}$ $r$-band luminosity $L_r$ calculated using $M_{r,\sun}=4.67$ as per \citet{Bell+03}.}
\end{deluxetable*}
%Typical uncertainties: $\sigma_\mathrm{R_{eff}}\sim0.04\arcsec$; \sersic{} $\sigma_n \sim 0.02$.

\begin{figure*}
{\centering
\noindent{\includegraphics[width=1.5\columnwidth]{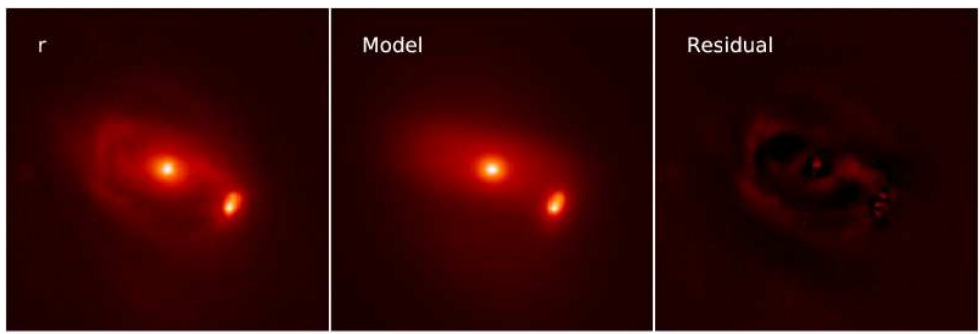}}
\caption{\textsc{galfit} model of the Was~49 system.  The model and residual images are on the same scale and (arcsinh) stretch as their respective input images.}
\label{fig:galfit}
}
\end{figure*}

Both \gala{} and \galb{} also have bulge-like components with \sersic{} indices of 1.43 and 1.07, respectively. These \sersic{} indices suggest that these components may be pseudobulges; however, further investigation is needed in order to confirm this classification.  An extended structure extending roughly north to south is also associated with \gala{}, which we modeled with a \sersic{} profile with index = 0.36, and which we interpret as being a tidal feature.  We also used a component to model a knot of emission associated with the extensive ionization to the north west of \galb{} that is likely residual emission line contamination in the $r$ band.  The fact that our morphological analysis does not find a disk component associated with \galb{} suggests that it should be classified as a dwarf elliptical (dE) galaxy \citep[e.g.,][]{Ryden+99}. The results of our \textsc{galfit} morphological fitting are shown in Figure~\ref{fig:galfit}.

In order to calculate the stellar bulge mass of \galb{}, we used the tight relationship between $g-r$ and the mass-to-light ratio from \citet{Bell+03}, which has a scatter of 20-50\%.  Because the $g$-band image is heavily contaminated by line emission, we have not attempted to model the $g$-band image with \textsc{galfit} and estimate $g-r$ directly.  Instead, we utilized the fact that $g-r$ is correlated with galaxy morphology, as quantified by the \sersic{} index $n$ \citep[e.g.,][]{BlantonMoustakas09}.  For galaxies with a \sersic{} index of $\sim1$, the typical $g-r$ color is about 0.4-0.5.  We estimated that the scatter on this relationship is $\sim0.2$ using galaxies from the bulge+disk decomposition catalog of \citet{Simard+11} with \sersic{} indices between 0.5 and 1.5.  Propagating this scatter and conservatively assuming an uncertainty of 50\% for the corresponding mass-to-light ratio $\log{(M/L_g)}=0.26$, we find that the stellar mass of the bulge of \galb{} is $\log{(M_\star/M_\sun)}=9.75\pm0.27$.

We also attempted to constrain the velocity dispersion of \galb{}, using a high S/N SDSS spectrum of a K-type star as a template for a faint \caii{}~K line seen in the BOSS spectrum.  We subtracted a second-order polynomial from the continuum on either side of the \caii{}~K line, and normalized the line to the corresponding feature in the rest-frame spectrum of \galb{}.  We iteratively stepped through a range of velocity dispersions $\sigma_\star$ between 70~km~s$^{-1}$ and 500~km~s$^{-1}$, in increments of 10~km~s$^{-1}$, convolving the star's \caii{}~K line with the corresponding Gaussian kernel, and measuring $\chi^2$ versus $\sigma_\star$.  At each step, we also added Gaussian noise to the original spectrum of \galb{}, taken from the variance of the spectrum, in order to estimate the significance of our results.  We found that the faintness of the \caii{}~K line precluded us from constraining the velocity dispersion of \galb{} to any useful range.

The unresolved nuclear component in \galb{} may be stellar in nature.  While its $g-r$ color of $\sim-0.1$ indicates considerable emission line contamination in the $g$ band, it has an $i-g$ color of $\sim-0.7$, much lower than the surrounding ionized region, which has an $i-g$ of $-0.4$ to $-0.1$.  Furthermore, we do not expect the optical counterpart of \galb{} to be composed significantly of direct emission from the AGN, owing to a high level of line-of-sight obscuration (\S\ref{X-rayResults}).  Using the empirical relationship between $E_{B-V}$ and $\nh{}$ for AGNs \citep{Maiolino+01}, $\nh{}=2.3\times10^{23}$~\cmsq{} corresponds to an $A_V$ of $\sim12$, given $R_V=3.1$.  If the nuclear component is stellar, then it may be a nuclear star cluster (NSC).  NSCs are massive, dense clusters of stars that are often present in late-type and bulgeless galaxies in the absence of a classical bulge, and are unique in that they are characterized by recurrent star formation and younger stellar populations \citep{Walcher+06}.  If we suppose that this possible NSC has a similar mass-to-light ratio as the surrounding bulge, then it has a stellar mass of $\sim2\times10^9$~M$_\sun$, which is consistent with expectations for NSCs, which are generally 1 to 10 times the mass of their SMBHs \citep{Seth+08}.  However, we emphasize that this is a highly tentative estimate, as the mass-to-light ratio is not known and the contribution to the $r$ band from residual line contamination and scattered AGN continuum has not been calculated.

To estimate the stellar mass ratio of \galb{} to \gala{}, we calculated their $r$-band light ratio.  Without their nuclear components, the $r$-band light ratio of \galb{} to \gala{} is 1:7.  With the addition of the nuclear components, the $r$-band light ratio is 1:6; however, we again urge caution that the unresolved nuclear component in \galb{} may be considerably contaminated by non-stellar emission.  We have not included the tidal feature associated with \gala{} and modeled in our \textsc{galfit} analysis.  This tidal feature may be some unknown mixture of material from \gala{} and \galb{}, and so we cannot quantify its contribution to either.  If, however, this feature is material entirely with \gala{}, the $r$-band light ratio is 1:13, in line with the $K$-band light ratio calculated below.  Conversely, if this feature is material entirely from \galb{}, the light ratio is 1:1, implying that the Was~49 system a major merger.  There are strong reasons why we do not consider this to be the case, as we will discuss in \S\ref{sec:Discussion}. 

We note that residuals in Figure~\ref{fig:galfit}, right, can be interpreted as being a mixture of tidal streams and regions of star formation in the disk of \gala{}, as can be seen in the $gri$ image presented in Figure~\ref{fig:griX}, left.  It is plausible that at least some of the tidal streams seen in the residual image originate in the progenitor galaxy of \galb{}, perhaps in a disk that was tidally stripped earlier in the merger.  In order to get a sense for the maximum possible mass that we might be missing, we summed the entire residual image.  We find an absolute $r$-band magnitude of the residuals of -18.1, corresponding to $2.3\times10^9$~\MSol{}, assuming a similar mass-to-light ratio.  The maximum \textit{total} galaxy mass of \galb{} would therefore be $\sim8\times10^9$~\MSol{}.

\begin{figure*}
{\centering
\noindent{\includegraphics[width=1.5\columnwidth]{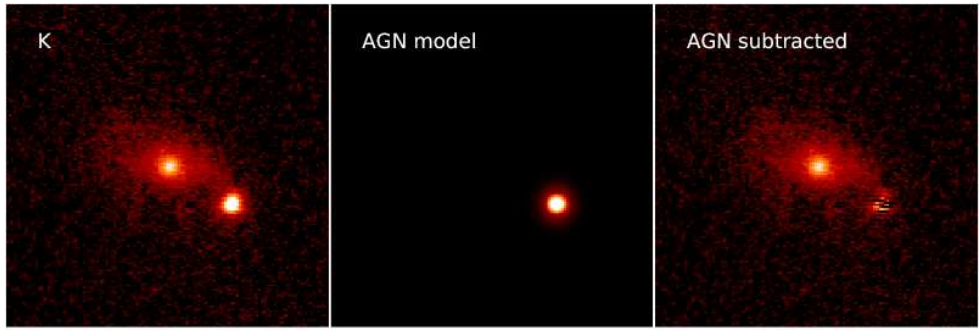}}
\caption{Result of our model of the AGN in \galb{} as an unresolved source in the $K$-band UKIDSS image, displayed with an arcsinh scale.}
\label{fig:UKIDSS}
}
\end{figure*}

We also note that we do not have reason to think that we may be underestimating the stellar mass of \galb{} due to dust obscuration.  We examined this possibility by measuring the $K$-band light ratio of \galb{} to \gala{} using an image from the UKIRT~Infrared~Deep~Sky~Survey \citep[UKIDSS;][]{Lawrence+07}, taken 2010~February~27.  At $2.2\micron$ the AGN is manifest as a dominant, unresolved source in \galb{} (Figure~\ref{fig:UKIDSS}, left).  Because of the dominance of the AGN and the shallowness of the UKIDSS image, it was difficult to model any extended emission in \galb{} using \textsc{galfit}, as was done for the DCT images.  We instead subtracted the AGN and measured the remaining emission by determining the expected $K$-band AGN luminosity given its intrinsic luminosity found in \S\ref{X-rayResults}.  Using five bright stars within a few arcminutes of Was~49, we made an empirical template of the PSF.  We found that we could accurately model the stars in the image with a PSF composed of three Gaussian components.  With our PSF template, we fit the unresolved source in \galb{}, allowing only the amplitude and position to vary (Figure~\ref{fig:UKIDSS}, middle).  We find a PSF $K$-band magnitude of 13.52, corresponding to $L_{K\mathrm{,observed}}=3.34\times10^{43}$~\ergs{}.  Before subtracting this unresolved source from the $K$-band image, we performed a check to ensure that it is consistent with the expected AGN emission. Using the value of $A_V\sim12$ for direct AGN continuum calculated above and the extinction curve of \citet{Cardelli+89}, we find $A_K\sim1.3$, implying an intrinsic AGN luminosity of  $L_K\sim10^{44}$~\ergs{}.   If we assume that the intrinsic spectral energy distribution is approximately flat in the near to mid-IR for luminous AGNs, then we can use $L_K\sim L_{6\micron}\sim\Lxtwoten$ \citep[e.g.,][]{Mateos+15} to calculate $L_{K\mathrm{,expected}}\sim4\times10^{43}$~\ergs{}, consistent with the intrinsic value, given the scatter on $E_{B-V}/\nh{}$ and $L_{6\micron}/\Lxtwoten$.  As the observed unresolved source in \galb{} is consistent with expectations for the AGN, we subtract it from the background-subtracted $K$-band image, and photometer the remaining emission in \galb{} for comparison with \gala{}.  We do not attempt to model any contribution to the $K$-band emission from the AGN in \gala{}, as the faint, soft X-ray source in \gala{} (\S\ref{X-rayResults}) suggests that the 2-10~keV luminosity, and therefore the expected $K$ band luminosity, is negligible.  We used the position and inclination angle for the disk of \gala{} (Table~\ref{tab:galfit}) to build an elliptical isophotal profile of \gala{}, which we interpolated across the position of \galb{} using a low-order polynomial and taking the median isophotal value.  Subtracting this profile, and estimating the variance from a source-free region in the UKIDSS image, the $K$-band light ratio is consistent with the Was~49 system being a $\sim$1:$15^{+17}_{-3}$ merger, where the confidence intervals are $\sim90\%$.  We emphasize that these confidence intervals assume that the AGN has been accurately subtracted from \galb{}.  This may not be the case, as there is scatter in the relationship between the intrinsic X-ray luminosity and the intrinsic IR luminosity of AGNs.  If our model of the AGN over-estimated the observed $K$-band luminosity of the AGN, then the $K$-band light ratio would be somewhat less.  It is less likely, however, that our model under-estimated the observed $K$-band luminosity of the AGN, as increasing the modeled luminosity would quickly lead to negative flux values when subtracting it from the $K$-band data.  With these considerations, we consider Was~49 to be a minor merger, with a mass ratio between $\sim$1:7 and $\sim$1:15.

% Was 49a "bulge" to disk ratio B/D = 0.48

\section{Discussion}
\label{sec:Discussion}

Our results show that the AGN in \galb{} is radiating at a very high luminosity ($\sim10^{45}$~\ergs{}), and is hosted by a possible dE galaxy in the disk of \gala{}.  Without the presence of a stellar host for \galb{}, \galb{} may have been a candidate for a recoiling black hole, as SMBH coalescence is thought to be the final stage in galaxy mergers and some SMBHs may be kicked out by gravitational-wave recoils at thousands of km~s$^{-1}$, manifesting as an offset quasar for up to tens of Myr \citep[e.g.,][]{BlechaLoeb08,Blecha+11}.  However, \gala{} would also have been left with no black hole, which is not the case, and the apparent co-rotation of \galb{} with the disk of \gala{} \citep{Moran+92} would be difficult to explain.  Compared to other known dual AGNs, the Was~49 system is highly unusual in that it is a minor merger with the more luminous AGN hosted in the secondary galaxy.  In all but one of the dual AGNs with BAT detections studied by \cite{Koss+12}, the BAT AGN is hosted in the more massive galaxy, and the sole exception (NGC~3758) is a major merger (stellar mass ratio of 1:2).  

We may gain some insight by comparing Was~49 to results from a numerical simulation of a gas-rich minor merger \citep{Callegari+11,VanWassenhove+12}.  This simulation of a coplanar, minor (1:10) merger bears some resemblance to the Was~49 system, given the apparent co-rotation of \galb{} within the disk of \gala{}, and the fact that the AGN in the primary galaxy is largely quiescent throughout the merger.  However, the secondary AGN only rarely reaches high luminosities ($L_\mathrm{bol}>10^{43}$~\ergs{}).  \citet{Capelo+15} also simulate a coplanar, 1:10 merger, and find similar results.  However, these simulations assumed that the central BHs were initially on the $M_\mathrm{BH}$-$M_\mathrm{bulge}$ relation: the \citet{HaeringRix04} relation in the case of former; the \citet{MarconiHunt03} relationship in the case of the latter.  Consequently, their secondary BHs start at masses of $6\times10^4$ and $3.5\times10^5$~\MSol, respectively, much smaller than the mass of $1.3\times10^8$~\MSol{} we find for \galb{}.  Assuming a similar Eddington ratio and scaling the bolometric luminosity of the secondary BH in the \citet{Capelo+15} simulation up by the ratio of BH masses, their simulation implies that \galb{} should have a bolometric luminosity of a few times $10^{44}-10^{45}$~\ergs{}, in line with the bolometric luminosity we calculated in \S\ref{X-rayResults}.  However, it is not clear how meaningful this scaling is, as the bulge mass of \galb{} is about 30~times larger.  Moreover, while the BHs in these simulations were $\sim0.03-0.2\%$ the mass of their host bulge as expected from scaling relations, the BH in \galb{} is apparently over-massive, at about 2.3\%.  It is therefore not certain how well these numerical studies can inform our picture of the Was~49 system.

% Callegari main disk galaxy:
% M_vir = 2.3e11 M_Sol
% M_d = 0.04*M_vir = 9.2e9
% M_b = 0.008*M_vir = 1.8e9

% Callegari satellite galaxy = main galaxy * 0.1 (M_b_satellite = 1.8e8)
% Callegari BH masses: 6e4, 6e5 (0.03% * M_b, M_b_satellite)

% Capelo 1:10: Use Marconi & Hunt (2003): MBH = 2e-3 * M_b
% G1: M_b = 1.77e9; MBH = 3.53e6
% G2: M_b = 0.18e9; MBH = 0.35e6

While the uncertainty on our SMBH mass of 0.5~dex means that it is possible that the SMBH is not as over-massive than the data suggests, we note that if \galb{} does follow the above BH-galaxy scaling relations, then the AGN would be radiating very near its Eddington limit.  However, the hard X-ray spectral index of $\Gamma=1.6$ implies that the AGN is radiating at a small fraction of its Eddington limit \citep[e.g.,][and references therein]{Brightman+16}.  For example, the relation between the Eddington ratio $\lambda_\mathrm{Edd}$ and $\Gamma$ from \citet{Brightman+13} implies that the AGN in \galb{} is radiating at only $\sim1\%$ of its Eddington limit, and other relations between $\lambda_\mathrm{Edd}$ and $\Gamma$ from the literature consistently imply that the Eddington ratio is only a few percent.  Given the bolometric luminosity calculated in \S\ref{X-rayResults}, the Eddington ratio for a $1.3\times10^8$~\MSol{} black hole is 0.08, meaning that our mass estimate is generally consistent with the X-ray spectral index.

Conversely, for a SMBH of $M_\textrm{BH}=1.3\times10^8$~\MSol{}, the host spheroid should have a mass of $\sim10^{11}$~\MSol{}, using the same scaling relations as \citet{Callegari+11} and \citet{Capelo+15}, making the Was~49 system a major merger, which is not the case, owing to two considerations.  First, in a major merger between two disk galaxies, both disks are severely disrupted, while in minor mergers the disk of primary galaxy is left relatively unperturbed \citep[e.g.,][]{Cox+08, Hopkins+09}, and there is no known mechanism by which only one member of an equal-mass system can be stripped.  Second, while a stellar disk can reform following a major merger \citep[e.g.,][]{Robertson+06,Governato+09}, the Was~49 merger is still ongoing, as \galb{} is at a projected separation from \gala{} of $\sim8$~kpc.  This implies that the disk of \gala{} is the original.  Moreover, \gala{} has a typical disk galaxy rotation curve \citep{Moran+92}, which strongly disfavors significant perturbation by \galb{}.

Empirically, Was~49 is not the only minor merger in which the smaller galaxy hosts the more luminous AGN.  For example, NGC~3341 is a minor merger composed of two dwarf galaxies merging with a giant disk galaxy.  One of the dwarf galaxies, 1/25 the size of the primary galaxy, hosts an AGN with an X-ray luminosity of $4.6\times10^{41}$~\ergs{}, while the primary galaxy is likely quiescent \citep{Bianchi+13}. \galb{} is 100 times more luminous, making it a unique system that may provide insight into the nature of minor-merger driven AGN fueling.

As noted in \S\ref{subsec:Morphology}, \gala{} is a consistent with being a pseudobulge galaxy, a morphology with a distinctly different and largely merger-free origin, suggesting that it has not gone through any major mergers in the recent past.  Indeed, the Was~49 system is isolated: a manual inspection of SDSS images/spectra shows that there are no other major galaxies ($M_r\lesssim-20$) within $\pm$1,000 km~s$^{-1}$ with a projected distance closer than about 1~Mpc.  \galb{}, being consistent with a dE galaxy in terms of mass and light profile, may have once been a late-type disk galaxy that was transformed into an elliptical morphology via galaxy `harassment' \citep[e.g.,][]{Moore+96}, or it may have been a primordial tidal dwarf galaxy \citep{DabringhausenKroupa13}.  The isolation of the system, however, implies that whatever morphological changes \galb{} underwent happened during the beginning of its encounter with \gala{}, and so it has not been severely tidally stripped \citep[unlike, for example, the SMBH-hosting ultracompact dwarf galaxy M60-UCD1:][]{Seth+14}, suggesting that the SMBH was intrinsically over-massive, or that perhaps the black hole's growth during the early phase of the merger happened well before the buildup of its host galaxy \citep[e.g.,][]{Medling+15}.  If \galb{} was originally a late-type/dwarf galaxy, its SMBH is a factor of $10^2$-$10^4$ times more massive than other black holes found in this galaxy type, which are typically between $10^4$-$10^6$~\MSol{} \citep[e.g.,][]{FilippenkoHo03, Barth+04, Shields+08, IzotovThuan08, Reines+11, Secrest+12, Dong+12, Reines+13, Secrest+13, Maksym+14, Moran+14, Secrest+15, Mezcua+16,Satyapal+16}, potentially giving new insight into how SMBHs form and grow in isolated systems.  For example, there has been recent work that has suggested that BH mass growth at higher redshifts precedes bulge growth \citep[e.g.,][]{Zhang+12}, while other work has found no such effect \citep[e.g.,][]{SchulzeWisotzki14}.  Numerical simulations suggest an evolution in BH/galaxy scaling relations with redshift \citep[e.g.,][]{Sijacki+15, Volonteri+16}, although it is not a dramatic effect, and numerical simulations also predict increasing scatter in BH/bulge scaling relations with decreasing bulge mass \citep[e.g.,][]{JahnkeMaccio11}.  Nonetheless, if early BH growth is indeed common, it may help explain systems like \galb{}.   A detailed discussion of this topic is, however, beyond the scope of this work.

\section{Conclusions}
We have performed a morphological decomposition of the Was~49 system using \textsc{galfit} and high-resolution optical images from the DCT, as well as a comprehensive X-ray analysis of \galb{} between 0.5-195~keV using data from \chandra{}, \nustar{}, and \swift{}. Our main results can be summarized as follows:

\begin{enumerate}
\item{Was~49, an isolated, dual AGN system, is a pseudobulge disk galaxy \gala{} in a minor merger ($\sim$1:7 to $\sim$1:15) with a potential dwarf elliptical galaxy \galb{} of stellar mass $5.6^{+4.9}_{-2.6}\times10^9$~\MSol{}.  The black hole mass of \galb{} is $1.3^{+2.9}_{-0.9}\times10^8$~\MSol{}, $\sim$2.3\% as massive as the galaxy it resides in and larger than black hole scaling relations predict.}

\item{The AGN in \galb{} is extremely luminous, with an intrinsic 0.5-195~keV luminosity of $\Lxfull=(2.4\pm0.2)\times10^{44}$~\ergs{}, and a bolometric luminosity of $L_\textrm{bol}\sim2\times10^{45}$~\ergs{}.  This is highly unusual for an AGN in the smaller galaxy of a minor-merger system, and makes Was~49 a unique system that can potentially yield insights into how AGNs are triggered in minor mergers.}

\end{enumerate}

\acknowledgements

We thank the anonymous referee for their thorough review of our manuscript.  We also thank Shai Kaspi (Tel Aviv University) for his advice regarding to the $\Lxtwoten$-$R_\mathrm{BLR}$ relation, Ren\'{e} Andrae (Max-Planck-Institut f\"{u}r Astronomie) for his helpful discussion of statistical methods, and Sara Ellison (University of Victoria) for her helpful input during the completion of this work.  Finally, we are indebted to Teznie Pugh and Jason Sanborn for their invaluable guidance during our observing run at the DCT.

This research has made use of the NuSTAR Data Analysis Software (NuSTARDAS) jointly developed by the ASI Science Data Center (ASDC, Italy) and the California Institute of Technology (Caltech, USA), as well as the UK Swift Science Data Centre at the University of Leicester. Funding for SDSS-III has been provided by the Alfred P. Sloan Foundation, the Participating Institutions, the National Science Foundation, and the U.S. Department of Energy Office of Science. The SDSS-III web site is http://www.sdss3.org/. SDSS-III is managed by the Astrophysical Research Consortium for the Participating Institutions of the SDSS-III Collaboration including the University of Arizona, the Brazilian Participation Group, Brookhaven National Laboratory, Carnegie Mellon University, University of Florida, the French Participation Group, the German Participation Group, Harvard University, the Instituto de Astrofisica de Canarias, the Michigan State/Notre Dame/JINA Participation Group, Johns Hopkins University, Lawrence Berkeley National Laboratory, Max Planck Institute for Astrophysics, Max Planck Institute for Extraterrestrial Physics, New Mexico State University, New York University, Ohio State University, Pennsylvania State University, University of Portsmouth, Princeton University, the Spanish Participation Group, University of Tokyo, University of Utah, Vanderbilt University, University of Virginia, University of Washington, and Yale University.  The UKIDSS project is defined in \citet{Lawrence+07}.  UKIDSS uses the UKIRT Wide~Field~Camera \citep[WFCAM;][]{Casali+07}.  The photometric system is described in \citet{Hewett+06}, and the calibration is described in \citet{Hodgkin+09}.  The WFCAM science archive is described in \citet{Hambly+08}.  This research made use of Astropy, a community-developed core Python package for Astronomy \citep{Astropy13}, and APLpy, an open-source plotting package for Python hosted at \url{http://aplpy.github.com}.  This research has made use of the NASA/IPAC Extragalactic Database (NED), which is operated by the Jet Propulsion Laboratory, California Institute of Technology, under contract with the National Aeronautics and Space Administration.

This research was performed while the author held an NRC Research Associateship award at Naval Research Laboratory.  Basic research in astronomy at the Naval Research Laboratory is funded by the Office of Naval Research.

\end{document}